**The Displacement Field Associated with the Freezing of a Melt and its Role in Determining Crystal Growth Kinetics**

Gang Sun, Alexander Hawken and Peter Harrowell[*]

*School of Chemistry, University of Sydney, Sydney NSW 2006 Australia*

**The atomic displacements associated with the freezing of metals and salts are calculated by treating crystal growth as an assignment problem through the use of an optimal transport algorithm. Converting these displacements into time scales based on the dynamics of the bulk liquid, we show that we can predict the activation energy for crystal growth rates, including activation energies significantly smaller than those for atomic diffusion in the liquid. The exception to this success, pure metals that freeze into face centred cubic crystals with little to no activation energy, are discussed. The atomic displacements generated by the assignment algorithm allows us to quantify the key roles of crystal structure and liquid caging length in determining the temperature dependence of crystal growth kinetics.**

Statement of Significance

*We demonstrate that an accurate estimation of the displacements associated with the transformation of liquid into crystal is necessary to explain the striking variations in the temperature dependence of the addition rate of liquid atoms to the growing crystal interface during freezing. An assignment algorithm, adapted from operations theory, is shown to provide a good estimate of these atomic displacements. As the assignment algorithm requires*

*only initial and target locations, it is applicable to all forms of structural transformations. In resolving a fundamental feature of the kinetics of freezing, a phenomenon of central importance to material fabrication, this paper also provides the tools to open new lines of research into the kinetics of structural transformation.*

Crystal growth from the melt is a common place phenomenon, one fundamental to material fabrication, whose kinetics depends on extraordinary sequences of cooperative events. In this paper we shall demonstrate how a map of this atomic choreography, the atomic displacement field, can be obtained from the structure of the interface alone and show how this displacement field allows us a conceptually deeper and, quantitatively, more accurate treatment of crystal growth kinetics. While the use of a displacement field that maps particles from liquid to crystal is novel, such maps are standard in the description of solid-solid transformations [1] where the existence of well-defined initial and final structures makes defining the associated transformation displacements a simple task. If the required displacements are less than a crystal layer spacing, the transition is classified as *diffusionless* or *displacive,* otherwise it is described as *diffusive* [1]. In freezing and melting transitions, the large excursions of atoms in a liquid obscure those movements essential for the transformation into the crystal phase. For this reason, freezing has not been previously analysed in terms of an atomic displacement field. The goal of this paper is to present an explicit definition and determination of the displacement field for the freezing of a liquid and to explain how to integrate this information into the theory of growth kinetics.

The 'classical' theory of crystal growth, as expressed in the theory of Wilson [2] and Frenkel [3] and the solid-on-solid model [4], considers crystallization to take place via the incoherent deposition of particles from an otherwise unstructured liquid into available sites on the growing crystal surface. The time scale for this deposition is assumed to be set by a liquid transport coefficient, diffusion [3] or viscosity [2]. This last assumption corresponds to an implicit proposition that crystal growth is diffusive in nature. This classical picture may be appropriate for precipitation from a dilute solution [5] but it ignores, in the context of growth from the melt, the strong local correlations that govern the structure and dynamics of the liquid state and the diffuse structure of the crystal-liquid interface [6,7]. These omissions are addressed, to a degree, in a Landau-Ginzburg treatment of freezing [8-10]. Eschewing reference to individual particles, this approach treats the crystal as a liquid with periodic density waves and the dynamics of crystal growth is now governed by the same time scale that governs density fluctuations in the liquid at the Bragg wave vector [8]. While the density wave representation makes sense for the construction of an equilibrium description of the freezing transition, it is not clear that the dynamics of ordering can be equally well described without reference to the motions of individual atoms. Our goal in this paper is to explore an alternative account of the process of freezing based on an extension of the existing formalism for solid-solid transformation.

Previously [11], having demonstrated that the crystal growth of a number of metals that form face centred cubic (FCC) crystals is non-diffusive, we argued that the absence of activated control arose as a consequence of the crystalline order that characterised the local groundstate of the liquid at the crystal-liquid interface. The evidence of the existence of these ordered interfacial groundstates is the propagation of the crystal interface during energy minimization. An example of this behaviour is shown in Fig. 1 along with an example of a system for which there is no advance of the interface. These interfacial ordered groundstates are in sharp contrast to those of the bulk liquid that exhibit no such order.

In a second publication [12], the mapping from liquid to crystal associated with energy minimization into the ordered interfacial groundstates was used to provide the displacement maps for the FCC metals and a molten salt, NaCl. An example of the resulting distribution of displacements for Cu is shown in Fig.2a. The time required for atoms in the bulk liquid to move the median value of the displacement field (see Fig. 2b) was found to provide a reasonable prediction of the observed activation energy for crystal growth.

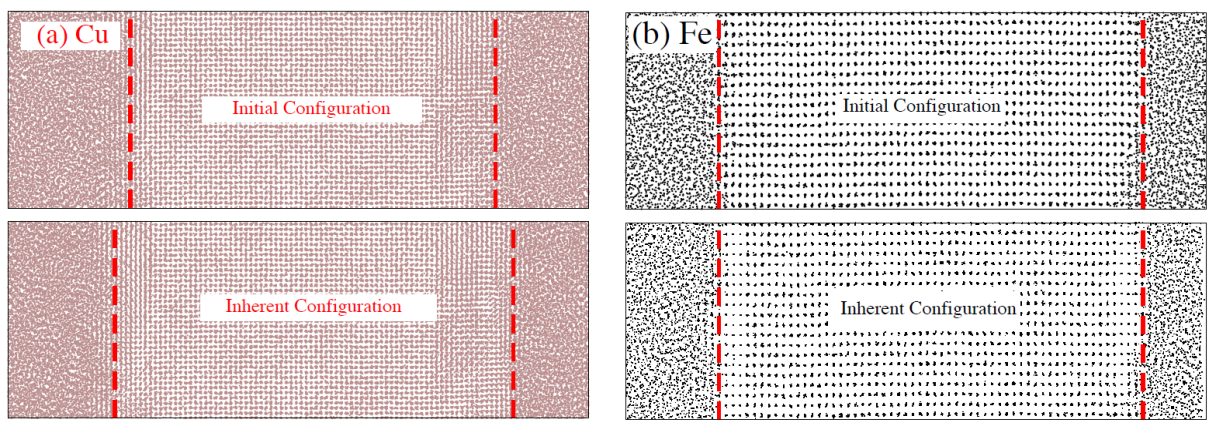

**Figure 1.** The change in a configuration corresponding to crystal and liquid as a result of the minimization of the potential energy for a) Cu (111) and b) Fe(110). The position of the interfaces in the initial and minimized configurations are indicated by vertical red dashed lines. Note that the Cu (111) interfaces advances while that for Fe does not.

The existence of ordered interfacial groundstates – the observation that formed the basis of the analysis in refs. [11,12] - is not the norm in crystal growth. In ref. [12] Fe and ZnS were shown to not order at the interface during energy minimization. The (111) face of a Lennard-Jones crystal is another example [11]. We propose that the majority of materials do not exhibit ordering on minimization. Clearly, we cannot simulate every system. Our claim is based on the generic experimental observations of activated control of crystal growth [14] for materials other than pure metals and some simple salts. Activation indicates an energy barrier for the ordering process at the interface and, hence, strongly *implies* the inaccessibility of the crystal from the liquid by any minimization algorithm that imposes a continuous decrease in the potential energy.

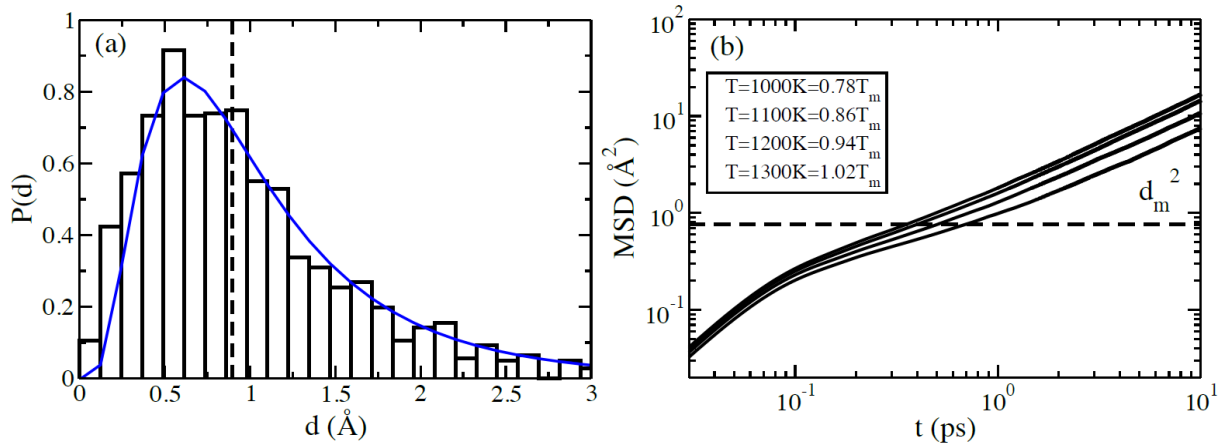

**Figure 2.** (a) The distribution of displacements associated with the crystal formation from the interfacial liquid atoms arising from the minimization of the potential energy in Cu (111) at T=1100K. The vertical dashed line indicates the median value $d_m$. The distribution can be well described by as log-normal distribution (blue curve) [13]. (b) Plots of the mean squared displacement MSD ($< \Delta r^2(t) >$) vs t in liquid Cu (111) showing the extraction of the T dependent time scale τ associated with the MSD $< d_m^2 >$. The different curves correspond to different temperatures with the larger value of MSD associated with a larger value of T (as indicated).

**Transformation Displacement Fields as a Problem of Optimal Transport.** If energy minimization does not provide a mapping from liquid to crystal at the interface, we must find another way of defining the transformation displacement field for an arbitrary crystal growth problem. In general, the displacement field corresponds to an assignment mapping where each liquid atom is assigned to a crystal site. The total number of distinct possible ways of assigning $N_L$ liquid atoms to $N_C$ crystal sites (both regarded as distinguishable by virtue of their locations) is $\dfrac{N_C!}{(N_C - N_L)!}$ where we have assumed that $N_C > N_L$ [15]. The displacement fields for most of these possible assignments will be of no physical significance. We propose that the set of displacements of interest is that with the *minimum average displacement*.

While this minimal field may not represent the path taken by any given ordering trajectory, it does provide a lower bound on the characteristic transformation displacement. This means that if this minimal length exceeds the threshold length obtained from the liquid cage length, then we can be confident that the associated freezing transition is diffusional in character.

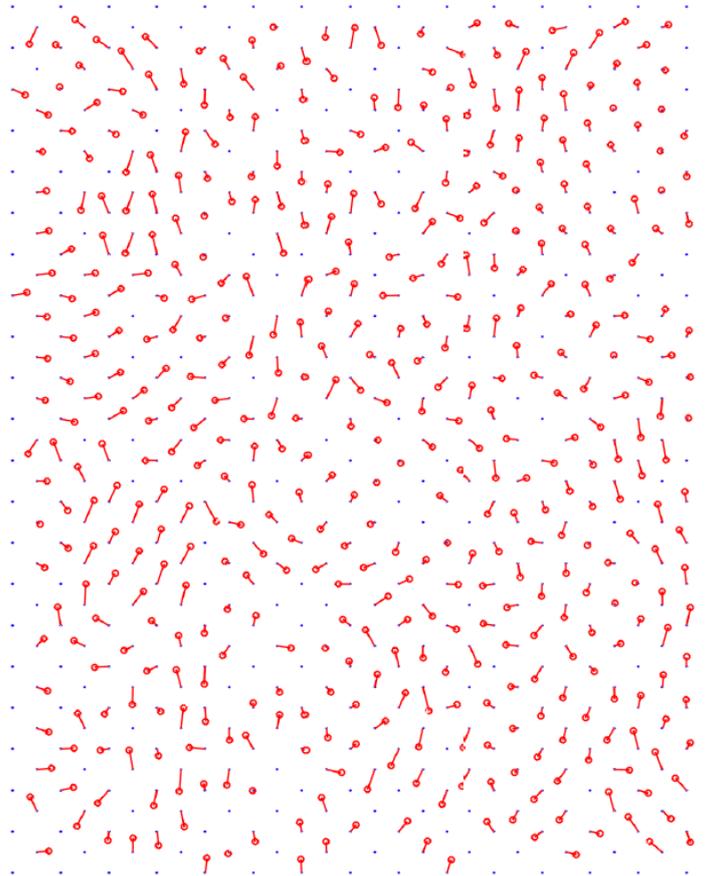

**Figure 3.** Map of the displacements due to the assignment analysis of Cu atoms in the interfacial plane (111). The red circles indicate the atoms position in the liquid and the lines indicate the displacement a nearby lattice site (blue small dots).

The problem of assigning atoms to crystal sites to minimize the mean displacement is an example of a problem well known in linear programming where it is called the optimal assignment problem or, more generally, the optimal transport problem [16]. Transport problems occur frequently in operations research – matching workers to tasks, exams to time slots, etc. A widely used solution is the Hungarian Algorithm (HA) described by Kuhn in 1955 [17]. We have implemented the HA for the crystal growth problem as follows. i) All calculations have been carried out with a particle-to-site ratio, $N_L/N_C \sim 0.6$. ii) The reported displacements are the components in the interfacial plane only. iii) The assignment is applied to atoms in the liquid that lies within two lattice spacings of the interface position. (A detailed explanation of these choices in the implementation of the HA is provided in the Supplementary Material.)   An example of the displacement field for an interfacial plane of the Cu crystal-liquid interface is shown in Fig.3. We note the presence of collective motions arising solely from the particle exclusion associated with the requirement of single occupancy of crystal sites.

**The Assignment Displacements and the Temperature Dependence of Crystal Growth Rates.**  The intrinsic crystal growth rate is the rate at which the crystal front propagates when unrestrained by heat or concentration diffusion. It is the growth rate controlled by the kinetics of microscopic ordering alone. The temperature dependence of these rates allows us to empirically distinguish diffusive from displacive transformations based on whether the associated activation energy is equal to or less than that for diffusion in the bulk liquid. We shall apply the standard resolution [18] of the growth rate v(T) into a thermodynamic term and a kinetic coefficient $k$(T) using

$$v(T) = k(T)\left[1 - \exp\left(\Delta\mu / k_B T\right)\right] \qquad (1)$$

where $\Delta\mu$ is the chemical potential difference and $k_B$ is the Bolzmann constant. Estimating the chemical potential difference as $\Delta\mu = (T - T_m)\dfrac{\Delta h}{T_m}$ ( with $T_m$ being the melting point and $\Delta h = h_{liq} - h_{cryst}$, with $h$ being the enthalpy per particle of the respective phase),  Eq. 1 allows us to extract the kinetic coefficient from the calculated rates v(T). In this paper we shall analyse the crystal growth kinetics of the following model systems. As examples of a metal and a salt whose crystal growth shows little to no activation barrier, we include Cu and NaCl. As examples of a metal and a salt whose crystal growth shows significant activation (and no sign of an ordered interfacial groundstate), we include Fe and ZnS. Finally, we include a family of binary Lennard-Jones equimolar mixtures that freeze into a cubic CsCl crystal. This mixture is based on a model due to Kob and Andersen [19] that has been studied extensively in the context of glass formation. These mixtures have been included as their crystal growth kinetics provide a continuous range of activation energies through a small adjustment of their interaction (as described below). The metals are modelled using an Embedded Atom Model (EAM) potential due to Foiles et al [20] and Mendelev et al [21], NaCl is modelled using the Tosi-Fumi potential [22] and ZnS is modelled using a potential due to Grünwald et al [23]. The binary Lennard-Jones mixtures $A_{50}B_{50}$ are modelled with the following interaction potential,

$$\phi_{ij}(r) = 4\varepsilon_{ij}\left[\left(\frac{\sigma_{ij}}{r}\right)^{12} - \left(\frac{\sigma_{ij}}{r}\right)^{6}\right] \tag{2}$$

where the masses are equal, $\varepsilon_{ij}$ and $\sigma_{ij}$ are the depth of the potential well between particle $i$ and $j$ and finite distance at which the potential (first) vanishes, respectively, $\varepsilon_{AA} = 1.0$ , $\varepsilon_{AB} = 1.5$ and $\varepsilon_{BB} = 0.5$ and $\sigma_{AA} = 1.0$ and $\sigma_{BB} = 0.88$ . The final parameter $\sigma_{AB}$ (which equals 0.8 in the original KA model) is allowed to vary over the range $0.76 \leq \sigma_{AB} \leq 0.84$ .

The crystal structures and melting points for these models are provided in Table 1. The crystal growth rates v(T) were calculated using molecular dynamics simulations from the LAMMPS [24] set of algorithms. As described previously [11,12], a uniform temperature is maintained throughout the run and the density is allowed to adjust through the inclusion of free liquid. These details, along with the crystal growth rates v(T) are provided in the Supplementary Material. For face centred crystals (FCC), we have calculated growth along the <111> direction, and for body-centered cubic (BCC) structure, the growth direction investigated is along the <110> direction. The growth directions of the rocksalt crystal for NaCl and the wurzite for ZnS are both along <100> direction. In the case of the CsCl crystal, we have considered growth along the <100> direction. Details of the simulation protocol along with the tabulated v(T) vs T data for all of the models is provided in the Supplementary Material.

In Table 1, we present the activation energy $E_a$, obtained empirically by fitting the rate coefficient $k$(T) (obtained from v(T) using Eq.1) ) to an Arrhenius function,

$$k(T) = k(T_m) \exp\left( -\frac{E_a}{k_B}\left[ \frac{1}{T} - \frac{1}{T_m} \right] \right)$$ , where $k(T_m)$ is the kinetics rate and $E_a$ is the energy

activation energy. For comparison, we also present the activation energy $E_D$ for self-diffusion in the homogeneous liquid, obtained by an Arrhenius fit to the self-diffusion coefficient D of the form $\ln D \propto -E_D / k_B T$. The generic signature of 'diffusionless' growth is $E_a / E_D$ significantly smaller than 1.0, a situation that describes the cases of Cu and NaCl. Fe and ZnS, in contrast, exhibit a growth coefficient with an activation energy similar to that of atomic diffusion in the liquid while the modified KA mixtures exhibit values over the range $0.41 \leq E_a / E_D \leq 0.91$. How do we account for this variety of degrees of activated control in crystal growth? We propose that the distribution of displacements generated by the optimal assignment algorithm, HA, provides the essential information we need. To translate the displacements into time scales we use the calculated mean squared displacement (MSD) of the bulk liquid as shown in Fig. 4 where a displacement d is translated into a time $\tau$ using the relation $d^2 = <\Delta r^2(\tau)>$. The temperature dependence of this relation arises from the temperature dependence of the liquid $<\Delta r^2(t)>$ (see Fig.4). Having generated a distribution of time scales P($\tau$) we calculate the average rate $<1/\tau>$ over this distribution and extract our *predicted* crystal growth activation energy $\tilde{E}_a$ by fitting the temperature dependence of $<1/\tau>$ to an Arrhenius temperature dependence (see Supplementary Material). The resulting values for $\tilde{E}_a$ are reported in Table 1.

| | *Crystal Structure* | $T_m$ | $k(T_m)$ | $E_D$ | $E_a$ | $\tilde{E}_a$ |
|---|---|---|---|---|---|---|
| **Cu** | FCC | 1275 K | 214(m/s) | 0.336(eV) | 0.04(eV) | 0.13(eV) |
| **NaCl** | rocksalt | 1074 K | 273(m/s) | 0.275(eV) | 0.078(eV) | 0.068(eV) |
| **Fe** | BCC | 1775 K | 1012(m/s) | 0.557(eV) | 0.358(eV) | 0.25(eV) |
| **ZnS** | wurtzite | 1750 K | 130(m/s) | 0.566(eV) | 0.403(eV) | 0.335(eV) |
| **AB ($\sigma_{AB}$= 0.76)** | CsCl | 0.687($k_B/\varepsilon_{AA}$) | 0.114[*] | 4.22($\varepsilon_{AA}$) | 3.84($\varepsilon_{AA}$) | 3.44($\varepsilon_{AA}$) |
| **AB ($\sigma_{AB}$= 0.80)** | CsCl | 0.772($k_B/\varepsilon_{AA}$) | 0.165[*] | 3.74($\varepsilon_{AA}$) | 1.53($\varepsilon_{AA}$) | 1.36($\varepsilon_{AA}$) |
| **AB ($\sigma_{AB}$= 0.82)** | CsCl | 0.77($k_B/\varepsilon_{AA}$) | 0.132[*] | 3.48($\varepsilon_{AA}$) | 1.74($\varepsilon_{AA}$) | 2.06($\varepsilon_{AA}$) |
| **AB ($\sigma_{AB}$= 0.84)** | CsCl | 0.752($k_B/\varepsilon_{AA}$) | 0.12[*] | 3.29($\varepsilon_{AA}$) | 2.45($\varepsilon_{AA}$) | 2.46($\varepsilon_{AA}$) |

**Table 1.** The crystal structure, melting points $T_m$, and parameters associated with the kinetics of crystal growth and liquid diffusion for the model systems. Presented are $k(T_m)$, the crystal growth kinetic coefficients (see Eq. 1) at $T_m$; the activation energies $E_D$ and $E_a$, for diffusion

and $k$(T), respectively, and $\tilde{E}_a$, estimated value of the growth rate activation energy based on the assignment analysis. [*] The kinetic coefficients $k(T_m)$ for the AB mixture are in units $\sqrt{\varepsilon_{AA}/m}$. The crystal face studies here are Cu (111), Fe (110), NaCl (100) and ZnS (100), and the (100) surface of the body centred CsCl crystal in the case of the Kob-Andersen (KA) mixtures.

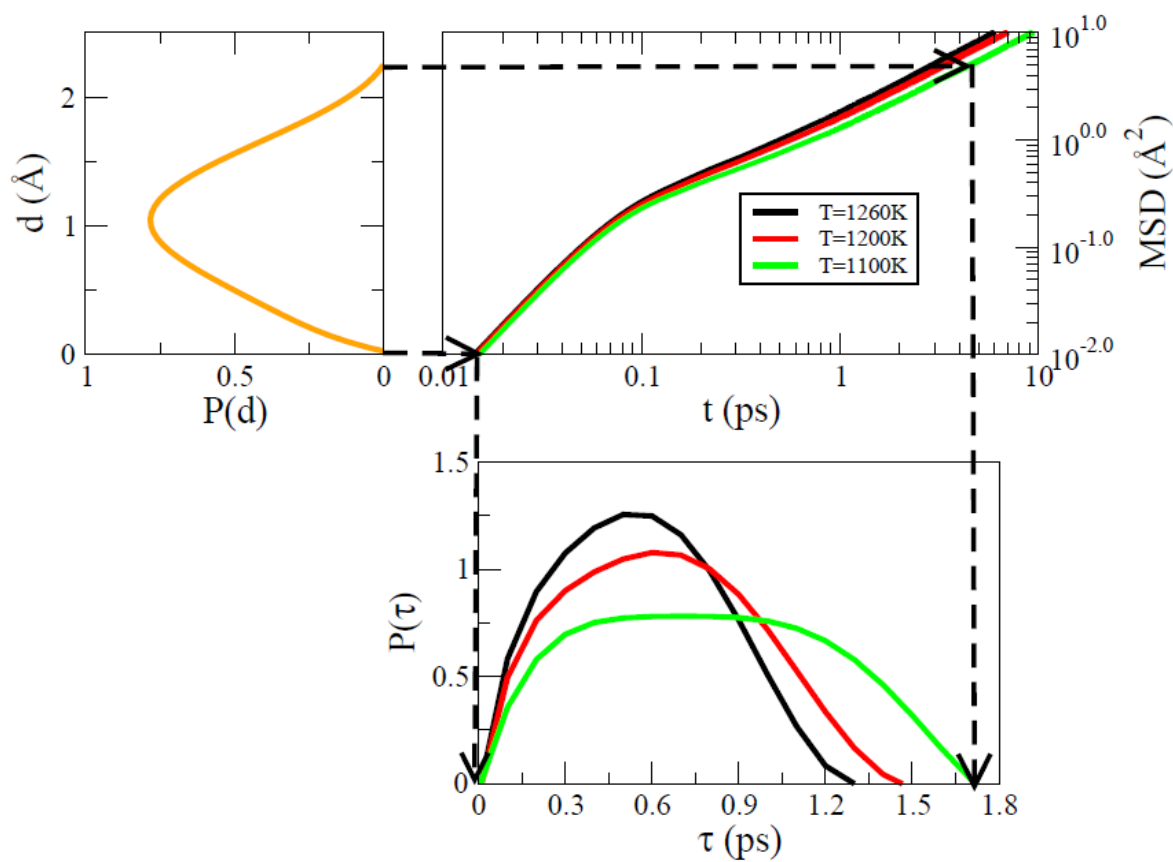

**Figure 4.** The mapping of the distribution of displacements (upper left) into the distribution of time scales (lower right) through the use of the time dependent MSD of the bulk liquid (upper right) at the temperature of interest as described in the text. The assignment displacements are for Cu (111) interface and the MSD data are for T = 1100K, 1200K and 1260K.

We find, as shown in Fig. 5, that the estimate of the activation energy $\tilde{E}_a$ for crystal growth based on the distribution of displacements generated by the assignment algorithm provides a reasonable prediction of $E_a/E_D$ and, hence, the degree of activated control of the crystal growth kinetics. A clear exception to this success is Cu where the assignment analysis results in a significant overestimate in the barrier to growth. It is likely that this represents a general limitation of the assigned displacements. When $E_a \ll 1.0$, the dynamics of growth is dominated by collective vibrations about the crystalline groundstate in the interface which is not adequately described by simply assigning atoms to sites based on minimizing the mean displacement. A more appropriate approach for this limiting behaviour is to use the characteristic vibrational frequency of the bulk crystal as proposed in ref. 11.

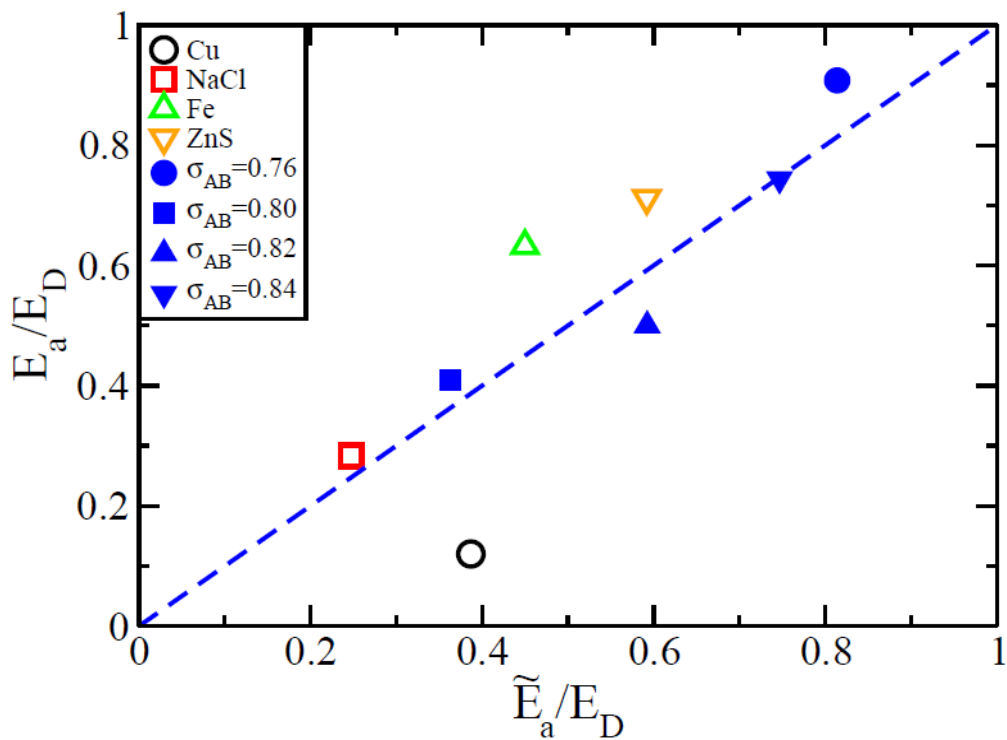

**Figure 5.** Plot of the simulated $E_a/E_D$ vs the predicted $\tilde{E}_a/E_D$ from the assignment analysis for Cu (111), Fe (110), NaCl (100) and ZnS (100), and the Kob-Andersen (KA) mixtures (100)

with modified parameters $\sigma_{AB}$ as indicated. The blue dashed line is $E_a / E_D = \tilde{E}_a / E_D$, provided as a guide to the eye.

The general success of the freezing displacement fields, as generated by the assignment analysis, in predicting the degree of activated control of crystal growth kinetics represents the major result of this paper. In using the liquid MSD data to map lengths to time scales we are following the same reasoning used in the earlier theories of the kinetic coefficient for crystal growth. Where we deviate from these earlier approaches in providing a more nuanced measure of the displacement lengths associated with ordering a liquid. The Wilson-Frenkel theory of crystal growth assumes that $E_a/E_D = 1.0$. The marked deviations from this 'classical' theory in Fig. 5 is the direct result of identifying length scales small enough to access dynamics that show little activated control. While the lengths we find using the optimal transport analysis may differ by only a small amount those assumed in the earlier studies [11,12], the difference can be physically significant and give rise to very large differences in both the temperature dependence of crystal growth rates as well as its absolute magnitude.

**Mapping to Alternate Crystal Structures: Polymorphs and Substitutionally Disordered Crystals.** We have reported that Cu freezes into an FCC crystal by a diffusionless process but the freezing of Fe to a BCC structure is diffusional. What is the origin of the difference? Is it some property of the liquid or does the difference lie with the different crystal structures? The optimal transport analysis provides a useful means of directly addressing question such as this since it allows us to select any target crystal structure we choose. If Fe was to freeze into the FCC structure, as opposed to BCC, how would the distribution of displacements and, hence, the nature of the transformation change? The assignment analysis allows us to address this question since we are free to select the target crystal to map the liquid particles to when generating the displacement distribution. The value of $\tilde{E}_a / E_D$ for Fe mapped to the FCC crystal is 0.46, essentially the same as the value, 0.45, when mapped to the BCC structure.

We can ask similar questions of the molten salts. In Table 2 we present the values of $\tilde{E}_a / E_D$ for the interfacial liquid NaCl and ZnS when mapped to rocksalt and wurzite structures, both compositionally ordered and compositionally random.

|  | rocksalt (random) | wurzite (random) |
|---|---|---|
| **NaCl** | 0.247(0.229) | 0.403(0.376) |
| **ZnS** | 0.605 (0.593) | 0.591(0.57) |

**Table 2.** The ratio $\tilde{E}_a / E_D$ for the assignment of the interfacial NaCl (100) and ZnS (100) to the rocksalt and wurzite structures. The lattice spacings of the target crystal structures were adjusted to match the cation-anion contact distance (see Supplementary Material). The values in brackets are for target crystals with random composition on the lattice sites.

Referring to Table 2, we find that NaCl does exhibit a significantly smaller value of $\tilde{E}_a / E_D$ when mapped to rocksalt than when mapped to wurzite. ZnS, on the other hand, behaves like Fe and shows little difference between the two possible crystal targets. We conclude that the manifestation of significant diffusive control, i.e. $E_a/E_D > 0.4$, does not depend on the crystal structure but is a consequence of the liquid. It appears that when the barrier to growth is particularly small (as in the case of NaCl), the choice of crystals may play a role in influencing the temperature dependence of growth. The results for the random crystals in Table 2 showed little variation for the values of $\tilde{E}_a / E_D$ for the compositionally ordered crystal. This is a surprising result as the observation of accelerated growth rates for disordered plastic crystal [25] would lead us to expect that disorder in the crystal would lead to faster kinetics. Clearly this is a question worth further study.

**Conclusions.** We have demonstrated that crystallization from the melt can occur via crystal growth with an activation energy that can take on a value $0 \leq E_a \leq E_D$ depending on the magnitude of the displacements associated with the transformation. We have introduced the Hungarian Algorithm, from optimal transport theory, to provide a general mapping of liquid atoms to crystal sites and so generate the transformation displacements between dense phases. We have shown in this paper, that these displacements, when translated into time scales using the dynamics of the bulk liquid, provide a good prediction of the degree of activated control of crystal growth. The crucial insight into understanding crystallization kinetics is that, when dealing with particle displacements on the order of the cage length of the liquid, very small changes in displacement can shift the dynamics from ballistic control to diffusive control. This means that understanding the kinetics of crystallization requires an accurate determination of the scale of movement needed for the transformation.

The exception to the success of the assignment analysis are those materials that exhibit very small or no activation barrier to crystal growth. The pure FCC-forming metals belong to this general group. As shown in Fig. 5 and Table 1, the assignment analysis significantly overestimates $E_a$ for Cu. Previously [11], we have argued that, in the absence of a barrier, the process of crystallization more closely resembles a complex collective phonon mode than the incoherent motion that characterises liquid dynamics. It is not surprising, then, that the simple minimization condition that defines the displacements generated by the Hungarian Algorithm would be a poor estimate of the coherent motions involved in ordering when the local potential energy minimum is crystalline. For these materials, the characteristic frequency of the crystal vibrations provides a better estimate of the magnitude of the ordering coefficient $k$.

Previous studies have observed anisotropy in crystal growth rates. In the case of the Lennard-Jones FCC crystal, the (111) surface exhibits activated growth while the (100) surface grows without any apparent barrier [26,27]. Applying the assignment analysis to these two surfaces, we find only a small decrease in the predicted activation energy for growth estimated for LJ(100) surface relative to that the LJ(111) surface, 1.6 $\varepsilon$ and 1.8 $\varepsilon$, respectively. This result for LJ(100) surface is consistent with the failure of the assignment analysis, already noted, when applied to barrierless growth such as that of Cu. We leave a more extensive study of the origin of anisotropy in crystal growth kinetics for future study.

Our introduction of transformation displacements and the HA algorithm for their calculation opens up new lines of study of crystallization kinetics based on measures of configurational proximity (i.e. the magnitude of the transformation displacements) between a disordered and ordered structure. As demonstrated here, a researcher has complete freedom in the choice of the target structure. Polymorphs, substitutional disorder or orientational disorder (in the case of molecules) could be considered in exploring how different aspects of structure influence the characteristic displacements atoms or molecules must undergo in a phase transformation. The optimal transport analysis can also be extended to transformation from one disordered structure to another in bulk liquids, quantifying the kinetic accessibility of different distinct configurations in a liquid. A number of these questions are the subject of current research.

**Acknowledgements**

PH would like to gratefully acknowledge a helpful conversation with Toby Hudson concerning the assignment problem and Paul Madden for discussions on solid state transformations. This work was supported by a Discovery Grant from the Australian Research Council. Data Availability Statement: All data discussed in the paper will be made available to readers.

Author Contributions: PH designed research, GS and AH performed research and analysed data and GS and PH wrote the paper.

* To whom the correspondence should be addressed: peter.harrowell@sydney.edu.au

**Supplementary Information**

*1. Simulation details and the orientational order parameter*

*2. The details of implementation of the Hungarian Algorithm (HA)*

*3. Assigning the lattice spacing to target crystal structures used in the HA*

*4. Tabulated crystal growth rate data and data for diffusion in liquids*

*5. Additional data on activation energy measurement for addition rate and the estimated from assignment.*

**1. Simulation details and the crystal order parameter**

*i) Initialization of crystal-liquid configuration.* The initial configurations contains $10\times10$ transformed unit cells in the x-y plane and 40 transformed unit cells in z direction for Cu, Fe, NaCl and ZnS systems, and 40 ($\sigma_{AA}$) × 40 ($\sigma_{AA}$) in x-y plane and 80 ($\sigma_{AA}$) along z direction for Kob-Anderson binary mixtures $A_{50}B_{50}$. The total particle number for Cu, Fe, NaCl and ZnS is about 64000 (depending on the crystal structure), for KA mixtures, is 163272. Then, for each system, the crystal particles of 10 unit cells in the middle were pinned, and the rest of the crystal was melted at a temperature above the melting point. Finally, we release the pinned particles and relaxed the whole crystal/liquid system for $3\times10^5$ timesteps at the melting temperature. These calculations were carried out using periodic boundary conditions in the three spatial directions and in the NPT ensembles. The pressure was held at zero and controlled by independently adjusting the cell dimensions in all three directions.

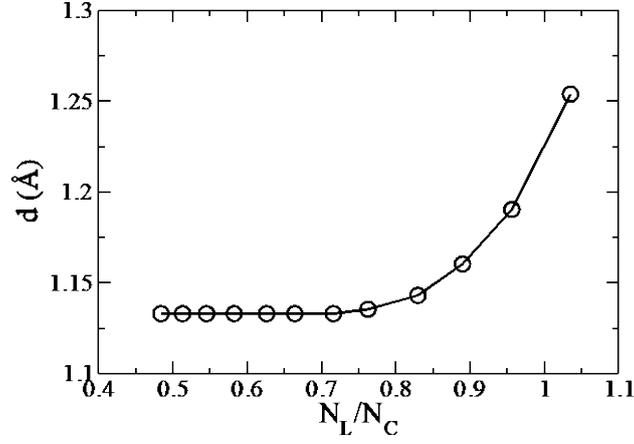

**Figure S1.** The median displacement values by Hungarian Algorithm (HA) as function of ratio of particles ($N_L$) and sites ($N_C$) for Cu. As the ratio goes above 0.6, $N_C \sim 1.67 \times N_L$, the mean displacement reaches an asymptotic value.

*ii) Crystal growth calculations.* During crystal growth, the diagonal components of the stress tensor, i.e. the pressures $P_{xx}$, $P_{yy}$ and $P_{zz}$ were independently maintained a zero using a Nose-Hoover barostat [1]. This allows for the system to adjust to the density change on crystallization along the z direction normal to the interface. To ensure a uniform temperature during crystal growth, we divided the system into a set of narrow slabs, each 5 diameters across, and independently thermostated each slab using a Berendsen thermostat [1].

*iii) Measuring the crystal growth rate.* To detect the crystal front, we use the distribution of the order parameter $\bar{Q}_{6,i}$ [2,3], which the average form of the local bond order parameter over all its neighbours and itself [3], which is defined as

$$\bar{Q}_{6,i} = \sqrt{\frac{4\pi}{13} \sum_{m=-6}^{6} \left| \bar{q}_{6m}(i) \right|^2}$$

(S1)

where

$$\bar{q}_{6m}(i) = \sum_{f \in F(i)} \frac{A(f)}{A} Y_{6m}(\theta_f, \varphi_f)$$

(S2)

where $A(f)$ is the surface area of the Voronoi cell facet $f$ separating the particle $i$ and its neighbouring particle that correspond to a given bond, and $A = \sum_{f \in F(i)} A(f)$ is the total surface area of the Voronoi cell boundary $F(i)$ of particle $i$, $Y_{6m}$ is the spherical harmonics, $\theta_f$ and $\varphi_f$ are the spherical angles of the outer normal vector $\mathbf{n}_f$ of facet $f$. We measure the crystal growth rate by detecting the width of crystal. The particle with $\bar{Q}_{6,i} > 0.95 \times \bar{Q}_{6,i}|_{crystal, T=T_m}$ is considered as a crystal. The energy minimization of the crystal/liquid interface was performed using the conjugate algorithm. During energy minimization, the crystal particles were fixed and the free boundary condition was applied along the crystal growth direction (z direction).

## 2. The details of implementation of the Hungarian Algorithm (HA)

*i) The choice of the particle-to-site ratio $N_L/N_C \sim 0.6$.* Given $N_L$ particles to be assigned, what number of crystal sites, $N_C$, is optimal? We find that on increasing $N_C$, the number of sites, the mean displacement d moved by particles as a result of the assignment decreases. This trend is shown in Fig. S1. It is clear in Fig.S1 that we reach an asymptotic value of $d$ at $N_L/N_C \sim 0.6$, and so have used this ratio in all the calculations reported here. *ii) The assignment displacement along different directions in space.* The presence of the interface breaks the symmetry of the system. This broken symmetry is evident in the distribution of displacements normal to the interface, i.e. $d_z$ and in the interfacial plane $d_{x,y}$. This difference between the two distributions is shown in Fig. S2. The normal displacement $d_z$ includes a component associated with net flow of material towards the surface as a result of the density change. As this flow is a) not directly associated with the ordering processes and b) unlikely to be governed by incoherent single particle motion, we have chosen to omit it and only use the transverse component of the assignment displacement, i.e. $d_{xy}$. This displacement is reduced in magnitude simply by virtue of being restricted to 2D, to correct for this, the final displacement is scaled by the geometric factor so that $d = d_{xy} \times \sqrt{\dfrac{3}{2}}$ .

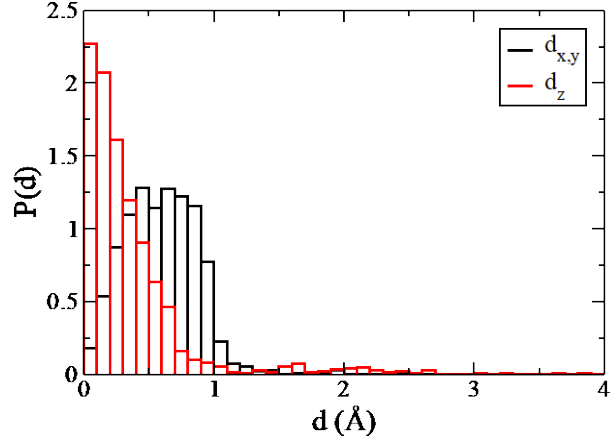

**Figure S2.** The distribution of displacements along in-plane ($d_{x,y}$) and the normal directions ($d_z$), associated with the interface liquid atoms in Cu (111) at T=1100K.

*iii) The assignment is applied to atoms in the liquid that lies within two lattice spacings of the interface position.* A key result of our earlier studies of the displacement fields [4,5] was the observation that the displacement field into the crystal from the interfacial liquid could be considerably smaller than the value from the bulk liquid. As it is this interfacial liquid adjacent to the crystal that is most directly associated with growth kinetics, we shall restrict our assignment mapping of liquid particles to a layer of liquid in the interface. The interface position is defined as follows. The order parameter profile of the interface $\overline{Q}_6(z)$ can be described by

$$\overline{Q}_6(z) = \frac{(\overline{Q}_{6,crystal} + \overline{Q}_{6,liquid})}{2} - \frac{(\overline{Q}_{6,crystal} - \overline{Q}_{6,liquid})}{2}\tanh\left(\frac{z - z_0}{w}\right) \qquad (S3)$$

where the $z_0$ is the interface position and w is the width of the interface. For Cu, the order parameter profile of the crystal-liquid interface is presented in Fig.S3. In general, the width of crystal-liquid interface is about 2-3 layers. The layer with position z=0 is defined as the first layer of the interface. In this work, we select particles that satisfy our condition of being a liquid, i.e. $\overline{Q}_6 < 0.3$, in the liquid slab defined by two lattice spacings from the interface position $z_0$.

### 3. Assigning the lattice spacing to target crystal structures used in the HA

The lattice spacing of the target crystal lattices are adjusted to match the characteristic length scale of the particle interactions. Specifically, for cubic lattices, the lattice constants are given by formula $a = 4 \times r / \sqrt{2}$ and $a = 4 \times r / \sqrt{3}$ for FCC and BCC respectively, where r is the atomic radius. The radius of Cu, Ni, Pb and Al are 1.278 Å, 1.245 Å, 1.75 Å and 1.409 Å, respectively. For the rocksalt lattice, the lattice constant is determined by the cation-anion distance, i.e. $a = 2(r_{cation} + r_{anion})$. For wurzite structure, the lattice constant is mainly determined the radius of anion, i.e. $a = 2\ r_{anion}$. The radius of $Cl^-$, $Na^+$, $S^{-2}$, and $Zn^{+2}$, are 1.81 Å, 1.01 Å, 1.91 Å, and 0.88 Å, respectively.

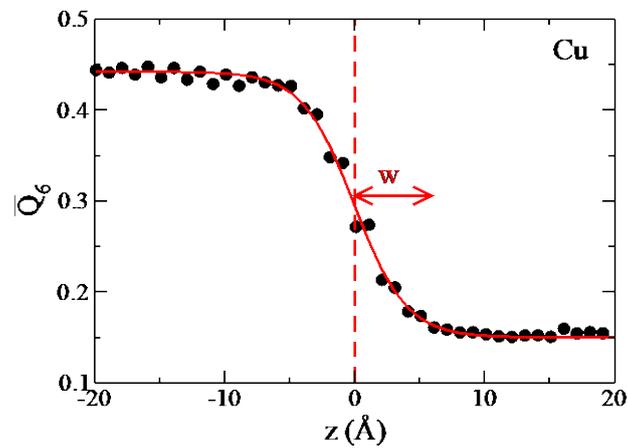

**Figure S3.** The order parameter profile of crystal-liquid interface for Cu (111). The red line is the fitted line based on Eq. S3. The interface position is $z_0=0$ and with w=4.16(Å).

**4. Tabulated crystal growth rate data and data for diffusion in liquids**

| Cu | | Fe | | NaCl | | ZnS | |
|---|---|---|---|---|---|---|---|
| T (K) | V(m/s) | T (K) | V(m/s) | T (K) | V(m/s) | T (K) | V(m/s) |
| 1275 | 0 | 1775 | 0 | 1074 | 0 | 1750 | 0 |
| 1260 | 8.0944 | 1760 | 12.011 | 1050 | 15.475 | 1700 | 7.554 |
| 1240 | 14.311 | 1740 | 22.86 | 1000 | 30.347 | 1650 | 8.979 |
| 1220 | 17.688 | 1720 | 29.43 | 950 | 52.488 | 1600 | 14.977 |
| 1200 | 22.319 | 1700 | 40.67 | 900 | 64.832 | 1550 | 17.211 |
| 1180 | 25.547 | 1680 | 49.63 | 850 | 82.44 | 1500 | 16.187 |
| 1160 | 29.083 | 1660 | 60.14 | 800 | 93.992 | 1450 | 17.764 |
| 1140 | 35.682 | 1640 | 64.67 | 750 | 106.362 | 1400 | 19.458 |
| 1120 | 36.1 | 1620 | 75.4 | 700 | 116.484 | 1350 | 19.849 |
| 1100 | 41.321 | 1600 | 80.5 | 650 | 123.351 | 1300 | 16.565 |
| 1080 | 45.772 | 1580 | 87 | | | 1250 | 17.435 |
| 1060 | 52.871 | 1560 | 96.3 | | | 1200 | 12.433 |
| 1040 | 60.925 | 1540 | 101 | | | | |
| 1020 | 58.867 | 1520 | 100.6 | | | | |
| 1000 | 68.294 | 1500 | 109.18 | | | | |
| 980 | 73.265 | 1460 | 115.7 | | | | |
| 960 | 74.2889 | 1400 | 123.6 | | | | |
| 940 | 74.9333 | 1300 | 130.8 | | | | |
| 920 | 80.6556 | 1200 | 133.8 | | | | |

**Table S1.** The crystal growth rates V(T) for Cu (111), Fe(110), NaCl(100) and ZnS(100).

| σ$_{AB}$=0.76 | | σ$_{AB}$=0.80 | | σ$_{AB}$=0.82 | | σ$_{AB}$=0.84 | |
|---|---|---|---|---|---|---|---|
| T | V | T | V | T | V | T | V |
| 0.68 | 0.00115 | 0.77 | 0.00069 | 0.77 | 0.00021 | 0.75 | 0.00013 |
| 0.66 | 0.00434 | 0.76 | 0.00306 | 0.768 | 0.00104 | 0.74 | 0.00346 |
| 0.64 | 0.00612 | 0.74 | 0.00932 | 0.76 | 0.00335 | 0.736 | 0.00450 |
| 0.62 | 0.00682 | 0.72 | 0.01435 | 0.74 | 0.00883 | 0.72 | 0.00777 |
| 0.60 | 0.00743 | 0.7 | 0.01849 | 0.72 | 0.01300 | 0.70 | 0.01191 |
| | | 0.68 | 0.02170 | 0.7 | 0.01820 | 0.68 | 0.01570 |
| | | 0.66 | 0.02678 | 0.68 | 0.02330 | 0.66 | 0.01963 |
| | | 0.64 | 0.03067 | 0.66 | 0.02901 | 0.64 | 0.02423 |
| | | 0.62 | 0.03408 | 0.64 | 0.03274 | 0.62 | 0.02686 |
| | | 0.60 | 0.03795 | 0.62 | 0.03700 | 0.60 | 0.03071 |
| | | | | 0.60 | 0.04024 | | |

**Table S2.** The crystal growth rates of KA binary mixtures (100) with different σ$_{AB}$. The units of temperature T and crystal growth rate V are $k_B/\varepsilon_{AA}$ and $\sqrt{\varepsilon_{AA}/m}$, respectively.

| Cu | | Fe | | NaCl | | ZnS | |
|------|------|------|--------|------|-------|------|-------|
| T | D | T | D | T | D | T | D |
| 1200 | 2.25 | 1760 | 0.3114 | 1070 | 7.466 | 1750 | 5.036 |
| 1150 | 2.1 | 1740 | 0.3 | 950 | 5.402 | 1700 | 4.608 |
| 1100 | 1.8 | 1720 | 0.285 | 900 | 4.54 | 1600 | 3.551 |
| 1050 | 1.56 | 1700 | 0.2717 | 850 | 3.769 | 1500 | 2.849 |
| 1000 | 1.25 | 1680 | 0.2641 | 800 | 2.956 | 1400 | 2.105 |
| 950 | 1.0 | 1660 | 0.2493 | 700 | 1.896 | 1250 | 1.238 |
| | | 1600 | 0.2138 | | | | |
| | | 1500 | 0.1629 | | | | |

**Table S3.** Diffusion Coefficients of Cu, Fe, NaCl, and ZnS. The units of T and D for Cu, Fe, NaCl and ZnS are K and $10^{-9}$m$^2$/s.

| $\sigma_{AB}$=0.76 | | $\sigma_{AB}$=0.80 | | $\sigma_{AB}$=0.82 | | $\sigma_{AB}$=0.84 | |
|------|---------|------|---------|------|---------|------|---------|
| T | D | T | D | T | D | T | D |
| 0.68 | 0.01402 | 0.76 | 0.03376 | 0.78 | 0.04261 | 0.76 | 0.04178 |
| 0.65 | 0.01039 | 0.74 | 0.02955 | 0.76 | 0.03735 | 0.74 | 0.03711 |
| 0.62 | 0.00751 | 0.72 | 0.02644 | 0.74 | 0.03309 | 0.72 | 0.03388 |
| 0.60 | 0.00581 | 0.7 | 0.02248 | 0.72 | 0.02941 | 0.70 | 0.02917 |
| 0.58 | 0.00436 | 0.68 | 0.01913 | 0.70 | 0.02544 | 0.68 | 0.02527 |
| 0.55 | 0.00259 | 0.66 | 0.01634 | 0.68 | 0.02186 | 0.66 | 0.02192 |
| 0.52 | 0.00134 | 0.64 | 0.01345 | 0.66 | 0.01919 | 0.64 | 0.01876 |
| | | 0.62 | 0.01122 | 0.64 | 0.01578 | 0.62 | 0.01561 |
| | | 0.60 | 0.00892 | 0.62 | 0.01309 | 0.60 | 0.01323 |

**Table S4.** Diffusion Coefficients of KA liquids with different $\sigma_{AB}$. The units of T and D for KA liquids, the T and D are presented with units, $k_B/\varepsilon_{AA}$ and $\sigma_{AA}\sqrt{\varepsilon_{AA}/m}$, respectively.

## 5. Graphical confirmation of Arrhenius temperature dependence of simulated k(T) and <1/τ> estimated from assignment.

Linear plots of ln (*rate*) vs 1/T are an empirical conformation that the rate can be described by an Arrhenius temperature dependence characterised by an activation energy Ea. In Fig. S4 we present these plots for both ln $k$(T) from crystal growth simulation and ln <1/τ>(T) from the assignment analysis as described in the text. We find linear relations in all cases with a slope proportional to the activation energy.

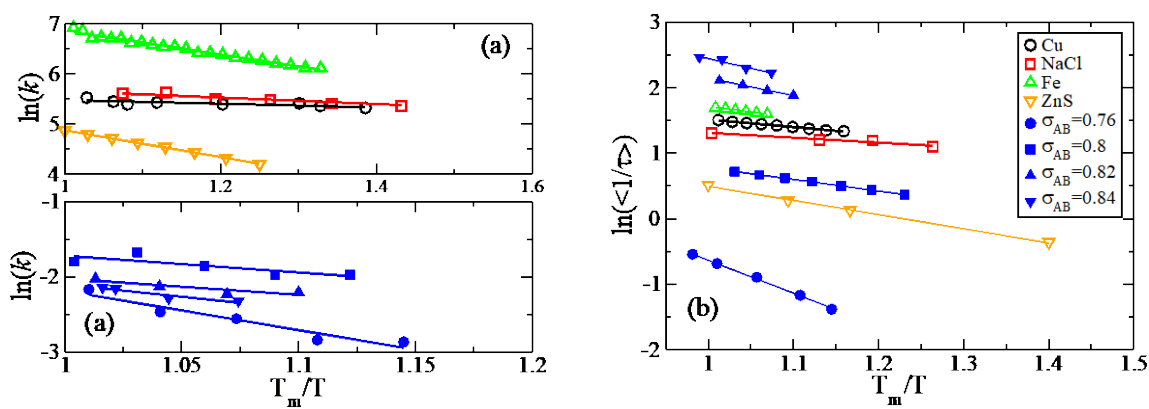

**Figure S4.** (a) Plots of lnk vs $T_m$/T for Cu(111), NaCl(100), Fe(110), ZnS(100) and KA model (B2(100)) with modified parameter $\sigma_{AB}$ . (b) Plots of ln<1/τ> transformed by assignment displacement accordingly, as function of vs $T_m$/T.  In each case, we can see that the lnk and ln<1/τ> reasonably follow the Arrhenius function, at least at T close to $T_m$.